\documentclass[twocolumn,pra,showpacs]{revtex4}
\usepackage{epsfig}
\usepackage{amsmath}
\usepackage{psfrag}
\usepackage{eufrak}
\usepackage{graphicx}
\usepackage{dsfont}
\usepackage{mathrsfs}
\usepackage[utf8]{inputenc}

\newcommand\ket[1]{\left|#1\right\rangle}
\newcommand\bra[1]{\left\langle#1\right|}

\newcommand\opone{\leavevmode\hbox{\small1\kern-3.8pt\normalsize1}}
\newcommand{\eps}{\mathcal E} \newcommand{\id}{\mathds 1}
\newcommand{\ten}{\otimes} \newcommand{\sigz}{\sigma_z}
\newcommand{\tr}{\mbox{tr}} \renewcommand{\rho}{\varrho}
\newcommand{\sys}{\mathcal S}
\newcommand{\res}{\mathcal R}
\newcommand{\rtot}{\rho_{\mbox{\tiny tot}}}

\begin{document}

\title{Quantum Decoherence of Two Qubits}
\author{Julius Helm}
\affiliation{Institut f\"ur Theoretische Physik, 
Technische Universit\"at Dresden, 
01062 Dresden, Germany}
\author{Walter T. Strunz}
\affiliation{Institut f\"ur Theoretische Physik, 
Technische Universit\"at Dresden, 
01062 Dresden, Germany}

\date{\today}
\begin{abstract}
  It is commonly stated that decoherence in open quantum systems is
  due to growing entanglement with an environment. In practice,
  however, surprisingly often decoherence may equally well be
  described by random unitary dynamics without invoking a quantum
  environment at all. For a single qubit, for instance, pure
  decoherence (or phase damping) is always of random unitary
  type. Here, we construct a simple example of true quantum
  decoherence of two qubits: we present a feasible phase damping
  channel of which we show that it cannot be understood in terms of
  random unitary dynamics. We give a very intuitive geometrical
  measure for the positive distance of our channel to the convex set
  of random unitary channels and find remarkable agreement with the
  so-called Birkhoff defect based on the norm of complete boundedness.
\end{abstract}

\pacs{03.65.Yz,03.67.Pp,03.65.-w}
\maketitle
\section{Introduction}
The loss of coherence in quantum systems is the hurdle that needs to
be overcome in attempts to make use of quantum mechanics on larger and
larger scales, most notably for quantum information tasks
\cite{NielsenChuang,Gisin}. Decoherence disentangles quantum states
\cite{Havel_NMR, Yu2003, Mintert}, which is why for experimental
realizations of quantum information processors it is of fundamental
importance to get a thorough understanding of the irreversible
processes involved \cite{NielsenChuang, Havel_NMR,
  Haeffner}. Decoherence is also put forward to explain the appearance
of classical properties in quantum systems \cite{ZurekRev, Giulini}.

In open quantum system dynamics, damping (population transfer) is to
be distinguished from decoherence (loss of phase relations in a
certain basis). While damping necessarily implies decoherence, the
converse need not be true for suitable interactions. Moreover,
decoherence often occurs faster than damping, so that a description of
the irreversible dynamics neglecting damping may be a valid approach
for a short enough time. A prominent example is the quantum Brownian
motion model \cite{CaldeiraLeggett}, where the damping term is
irrelevant during rapid decoherence \cite{ZurekRev, Giulini,
  HaakeStrunz}. Pure decoherence is usually referred to as {\it
  dephasing} or {\it phase damping}.

It is often stated that decoherence of an open quantum system is due
to growing entanglement between system and environment \cite{Zurek,
  SeligmanCompagnoToda}. Nevertheless, more often than one might
think, the resulting irreversible dynamics of the open system may be
modeled entirely without invoking a quantum environment. Rather, the
dynamics turns out to be indistinguishable from a random unitary (RU)
evolution, which can be thought of as originating from classical
fluctuations -- sometimes called ``random external fields''
\cite{AlickiLendi, NielsenChuang} (see also the corresponding
discussion in \cite{ZurekRev}). Note that in one of the most
detailed experimental studies of decoherence in ion traps, classical
fluctuations (i.e., RU dynamics) are used to cause controlled
decoherence \cite{Myatt,Libby}. In NMR decoherence studies, too,
fluctuating classical fields are employed \cite{Havel_NMR}.

There are many more relevant instances of decoherence that are of RU
type. In fact, for a single qubit or qutrit, any possible phase
damping is RU \cite{LandauStreater, DAriano}. Also, the very often
employed model of Markovian dephasing (Lindblad master equation)
\cite{Mintert, Werlang, Carvalho} belongs to this class since any
self-adjoint Lindblad operator may be identified with a white noise
term in a suitable Hamiltonian. In this vein, the quantum Brownian
motion master equation mentioned above (neglecting damping) follows
from a white noise force term in the Hamiltonian, therefore being
RU. Based on Feynman and Vernon's influence functional approach, one
sees that the latter is not even restricted to the usual high
temperature limit but can easily be extended to any temperature using
colored noise -- as long as times are short enough so that the
non-Markovian damping kernel may be neglected \cite{feynman,weiss}.

We conclude that many widely used decoherence scenarios are of the RU
class. Still, from the work of Landau and Streater
\cite{LandauStreater} -- which plays a central role for our results
here -- it is known that phase damping is not necessarily RU. However,
there is no known simple criterion able to decide whether a given
phase damping dynamics belongs to the RU class. From a more practical
point of view, a test for a channel to be of non-RU type is also of
relevance for quantum error correction \cite{ZurekRev} for it is known
that such errors may not be fully corrected \cite{GregorattiWerner}.

We deem it desirable to have a simple, explicit example of {\it
  quantum decoherence} at hand of which it is known that it cannot be
expressed using stochastic Hamiltonians.  Using a two qubit system we
present a model of which we show that phase damping truly rests on
growing entanglement with a quantum environment. Somewhat similar to
studies by Havel and co-workers, our proposal may be implemented in
NMR systems \cite{Havel_NMR}, and also in ion trap quantum computers
\cite{Haeffner}.

In a first step, we choose as environment a single, third qubit. The
proof that the corresponding decoherence cannot be understood in terms
of RU dynamics follows immediately from the work of Landau and
Streater (Sec. II). Remarkably, using the Bloch sphere picture, we
find a nice geometrical measure (a volume) that indicates how
``non-RU'' the dynamics is. This quantity correlates surprisingly well
with the distance of the quantum decoherence channel from the convex
set of RU dynamics (using the so-called cb-norm, Sec. III). In Sec. IV
we extend our model to include genuine irreversibility.

\section{Phase damping channels and extremality}
The dynamics of a quantum state $\rho\rightarrow \rho'=\eps[\rho] =
\sum_i K_i \rho K_i^\dagger$ is a completely positive map (or quantum
channel) with Kraus operators $K_i$ \cite{NielsenChuang} (neglecting
initial correlations). A decoherence or phase damping channel belongs
to the class of doubly stochastic channels. These are trace preserving
and unital (mapping the identity onto itself), corresponding to
$\sum_i K_i^\dagger K_i = \mathds 1$ and $\sum_i K_i K_i^\dagger =
\mathds 1$, respectively. The question about the nature of the
irreversibility (entanglement vs RU) is then in close analogy to the
classical Birkhoff theorem, stating that every real doubly stochastic
matrix can be written as a convex sum of permutations
\cite{LandauStreater}. In the quantum case, this corresponds to the
question of whether the set of doubly stochastic quantum channels is
identical to the set of RU channels $\eps [\rho] = \sum_i p_i U_i \rho
U_i^\dagger$ with unitary $U_i$, $p_i>0$, and $\sum_i p_i = 1$ (see
\cite{DAriano,RUpapers} for recent work on RU channels).

Decoherence or phase damping channels are among the simplest
conceivable maps. They are defined by the requirement that, in the
given basis $\{|n\rangle\}$ with $1\le n\le d$, no population transfer
takes place: $\langle n |\rho|n\rangle = $const.  The only effect of
the ``environment'' is thus to change coherences $\langle m
|\rho|n\rangle$ with $m\neq n$.  Thus, the Kraus operators have to be
diagonal in this basis, $K_i=$diag$(a_{i1}, a_{i2}, \ldots , a_{id})$
and, correspondingly, the whole map $\rho'=\eps[\rho]$ is diagonal,
\begin{equation}
  \label{dephasingchannel}
  \rho'_{mn} = \langle a_n|a_m\rangle\; \rho_{mn}
\end{equation}
with $\{|a_n\rangle =(a_{1n},a_{2n},\ldots,a_{rn})\}$ any set of $d$
normalized complex vectors \cite{HavelHadamard,Paulsen}.  Phase
damping channels are just the diagonal, doubly stochastic quantum
channels.

If the phase damping channel $\eps$ results from the coupling between
the system and a quantum mechanical environment, the vectors
$\ket{a_n}$ may be understood as relative quantum states of the
environment, relative to the states of the distinguished basis
\cite{GorinStrunz}. Yet, this need not be: for a single qubit the most
general phase damping channel
\begin{equation}
  \rho' = e^{-i\phi_0\sigma_z}\left( p \rho + (1-p)\sigma_z \rho \sigma_z 
  \right) e^{i\phi_0\sigma_z}
\end{equation}
may obviously be obtained from $U_\phi = e^{-i\phi\sigma_z}$ with a
random variable $\phi$ with
$\phi_0=\langle\!\langle\phi\rangle\!\rangle$ and
$p=\langle\!\langle\cos^2(\phi-\phi_0) \rangle\!\rangle$. Landau and
Streater show in \cite{LandauStreater} that for the case of $d\geq 4$,
e.g., for at least a two-qubit system, there exist non-unitary
extremal maps in the set of diagonal doubly stochastic quantum
channels: there are phase damping channels of two qubits that are not
of RU type.

\begin{figure}[t]
  \center
  \includegraphics[width=0.6\linewidth]{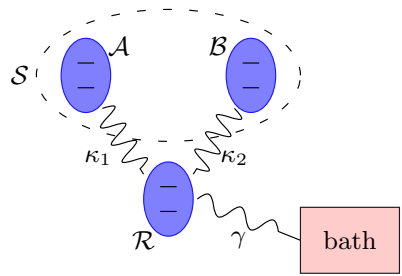}
  \caption{\small (Color online) Phase damping channel on the system
    $\sys$ of qubits $\mathcal A$ and $\mathcal B$. The
    ``environment'' initially consists of just a single qubit $\res$,
    which is later coupled to an additional zero temperature bath.}
  \label{fig:figure1}
\end{figure}

We set out to construct such a channel, which requires having a test
for a channel's extremality. A given diagonal doubly stochastic
channel $\eps$ is extremal iff it admits a Kraus representation
$\eps[\rho]=\sum_{i=1}^r K_i \rho K_i^\dagger$, where $\{K_i^\dagger
K_j\}_{i,j=1}^r$ is a linear independent set of matrices
\cite{LandauStreater}. This linear independence is equivalent to a
quality of the associated vectors $\{ \ket{a_1},\ldots, \ket{a_d} \}
\subset \mathds C^r$ called ``full set of vectors'' (FSOV)
\cite{LandauStreater}, which is attained if, for a complex matrix $M
\in \mathds C^{r\times r}$, $\langle a_n| M | a_n \rangle =
0 \; \forall \;n$ implies $M=0$. Note that in case of a two-qubit
channel extremality implies $r\leq2$: for $r=1$ this is just unitary
dynamics, so that $r=2$ gives the only possibility of an extremal,
non-unitary phase damping channel, implying the $\ket{a_n}$ to be
single-qubit states.

Based on this relation we give a simple test for extremality.  With
$\vec\Sigma:= (\id,\vec\sigma) =(\id,\sigma_x,\sigma_y,\sigma_z)$ we
denote the vector containing the usual basis of linear operators in
two-dimensional Hilbert space. Then, for the states $\ket{a_n}$ of the
environmental qubit, the Bloch representation reads
$\ket{a_n}\bra{a_n} = \frac{1}{2}(\mathds 1 + \vec b_n \cdot \vec
\sigma)=: \vec B_n \cdot \vec\Sigma$, where $\vec B_n =
\frac{1}{2}(1,\vec b_n)$. When also rewriting $M$ in this basis, $M =
\vec K \cdot \vec\Sigma$, with $\vec K \in \mathds C^4$, the FSOV
condition demands that $ \langle a_n | M |a_n \rangle = \vec B_n \cdot
\vec K = 0 $ for all $n$ implies $M = 0$ and, accordingly, $\vec K =
0$. Hence, the vectors $\{\vec B_1,\ldots,\vec B_4 \}$ have to be
linear independent, so that we get the following equivalence:
\begin{eqnarray}
  \begin{array}{c}
    \{\ket{a_1}, \ldots, \ket{a_4} \} \\[2mm]
    \mbox{is a FSOV} 
  \end{array}
  \Leftrightarrow 
  V_t := \frac{1}{6} \det
  \left(
    \begin{array}{cccc} 
      1 &\cdots& 1 \\{}\\
      \vec b_1 & \cdots & \vec b_4
    \end{array}
  \right) \ne 0. \quad
  \label{eq:volume}
\end{eqnarray}
In addition we arrive at a geometrical interpretation of the FSOV
condition: the channel is extremal iff the Bloch vectors $\vec b_n$ do
not point to the same hyperplane in $\mathds R^{3}$, or, equivalently,
iff the volume $V_t$ of the $3$-dimensional tetrahedron spanned by the
Bloch vectors is different from zero (see also fig.~\ref{fig:figure2}
(c) ). Note that we here discuss the case of a system of two qubits
only, the results can however be extended easily to arbitrary
dimension.

\section{Quantum Decoherence due to Single-Qubit ``Environment''}
After having presented the technical prerequisites we now want to
consider a simple model based on a two-qubit system $\sys$ (qubits
$\mathcal A$ and $\mathcal B$) interacting with a quantum mechanical
``environment,'' consisting of only one single qubit $\res$
(cf.~Fig.~\ref{fig:figure1}). Evolution of the compound system shall
be described by the Hamiltonian
\begin{eqnarray}
  \label{eq:hamiltonian}
  H = H_{\sys} + H_{\mathcal I} + H_{\res},  
\end{eqnarray}
where $H_{\sys} = \Omega_1 \sigz^{(\mathcal A)} + \Omega_2
\sigz^{(\mathcal B)}$ denotes the system Hamiltonian, $H_{\mathcal I}
= \kappa_1 \sigz^{(\mathcal A)} \sigz^{(\mathcal R)} +
\kappa_2 \sigz^{(\mathcal B)} \sigz^{(\res)}$ describes the
interaction between system and environment, and $H_{\res} = \vec
\Gamma \cdot \vec \sigma^{(\res)}$ ($\vec \Gamma =
(\Gamma_x,\Gamma_y,\Gamma_z)\,$) gives the free evolution of the
reservoir qubit.

With the usual product initial state $\rho \ten \sigma$,
for any given time $t$ the reduced dynamics of the two qubits
defines a phase damping channel
\begin{eqnarray}
  \label{eq:tqd-channel}
  \eps_t [\rho] =: \rho'= \tr_{\res} \left( e^{-i H t}
    \left( \rho \ten \sigma \right) e^{i H t} \right). 
\end{eqnarray}

\begin{figure}[t]
  \includegraphics[width=.9\linewidth]{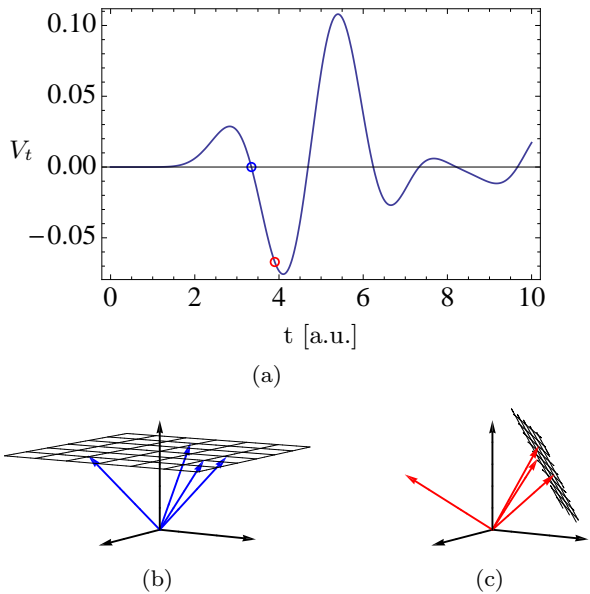}
  \caption{\small (Color online) (a) The volume $V_t$
    (\ref{eq:volume}) against time $t$ for a given set of parameters
    $\kappa_1, \kappa_2, \vec\Gamma$. For values of $V_t = 0$ [left
    circle and (b), exemplary] the corresponding Bloch vectors are
    coplanar, and the phase damping channel is RU. For $V_t \ne 0$
    [right circle and (c), exemplary] the dynamical vectors $\{
    |\psi_t^{(n)}\rangle \}$ form a FSOV, so that the corresponding
    Bloch vectors are not coplanar. In this case, the model gives a
    quantum decoherence channel.}
  \label{fig:figure2}
\end{figure}
Rewriting $H$ in the basis of the system's energy eigenstates, $H =
\sum_{n=1}^4 \ket n \bra n \ten \tilde {H}_{\res} ^{(n)}$, and taking
the initial state of the reservoir to be pure, i.e., $\sigma =
\ket{\psi_0}\bra{\psi_0}$, we get
\begin{eqnarray*}
  \label{eq:tqd-channel2}
  \rho'_{mn} = \tr \big( e^{-i \tilde{H}_{\res}^{(m)}t}\ket{\psi_0}
  \bra{\psi_0}
  e^{i \tilde{H}_{\res}^{(n)}t} \big) \rho_{mn} 
  = \langle \psi_t^{(n)}|\psi_t^{(m)} \rangle \rho_{mn}, 
\end{eqnarray*}
with the relative states of the environment $|\psi_t^{(n)}\rangle:=
e^{-i \tilde{H}_{\res}^{(n)}t}\ket{\psi_0}$, $n=1,\ldots,4$. The
extremality of the phase damping channel may now be checked by
calculating the volume $V_t$ (\ref{eq:volume}).  We find that
extremality requires the parameters of our model to meet essentially
three conditions:
\begin{itemize}
\item[(I)]
  Asymmetric coupling: $0 \ne \kappa_1 \ne \kappa_2 \ne 0$,
\item[(II)] 
  $\Gamma_x\ne 0$ or $\Gamma_y\ne 0$, and
\item[(III)]
  $\Gamma_z \ne 0$.
\end{itemize}
For $V_t=0$ the channel is not only non-extremal, we can further show
that it is also random unitary. First note that for a channel with
$\vec b_1,\ldots,\vec b_4$ pointing to a plane parallel to the
$x$-$y$-plane random unitarity follows immediately, for we can write
$|\psi_t^{(n)}\rangle = (\sqrt{1-p}\; e^{i\varphi^{(n)}_1}, \sqrt{p}
\; e^{i\varphi^{(n)}_2}) $ with the same $p$ for all $n=1,\ldots,4$,
resulting in the Kraus form $\eps[\rho] = (1-p) U_1 \rho U_1^\dagger +
p U_2 \rho U_2^\dagger$. For arbitrary coplanar Bloch vectors $\vec
b_1,\ldots,\vec b_4$ a suitable rotation of both the initial state
$|\psi_0 \rangle$ and the Hamiltonians $\tilde{H}_{\res}^{(n)}$ leaves
the phase damping channel unaltered, whereas the plane spanned by the
new Bloch vectors is again parallel to the $x$-$y$-plane.

\begin{figure}[t]
  \includegraphics[width=.9\linewidth]{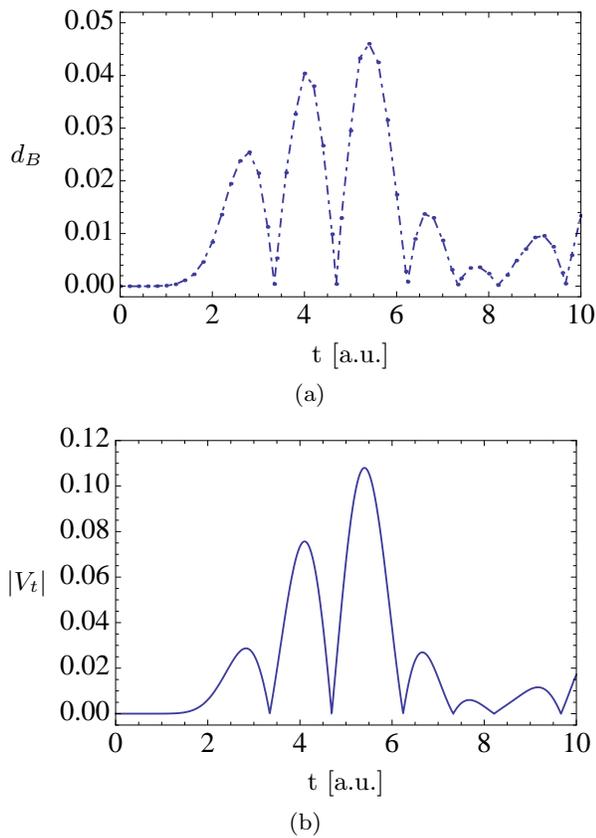}
  \caption{\small (Color online) (a) the calculated cb-norm distance
    $d_B$ of our quantum phase damping channel to the set of random
    unitary channels. The qualitative agreement with (b) the absolute
    value $|V_t|$ of the volume spanned by the Bloch vectors
    characterizing the channel is remarkable.}
  \label{fig:figure3}
\end{figure}

In Fig.~\ref{fig:figure2} (a) the volume $V_t$ is plotted as a
function of time $t$ for a realization with conditions (I) - (III)
met. Based on our considerations we can conclude that for almost all
times the corresponding phase damping channel $\eps_t$ is an extremal
channel ($V_t \ne 0$). From the decrease in Purity $P(\rho') =
\tr(\rho'^2)$, we can exclude unitary dynamics, assuring the channel
to be a genuine quantum decoherence channel
[cf. Fig. \ref{fig:figure4} (a), solid line].

We now want to quantify the ``quantumness'' of the decoherence by
determining the norm distance of the obtained channel to the set of RU
channels, also called the Birkhoff defect, $d_B$
\cite{WernerOpenProblems}. The norm distance is calculated in terms of
the cb-norm (norm of complete boundedness, for definitions and
properties see \cite{Johnston,Paulsen}). It involves numerical
minimization over (a) the convex set of RU channels, and (b)
equivalent operator sum representations of the channel occurring from
the difference of the given phase damping channel and the
corresponding random unitary channel. In order to find the global
minimum we use several starting points, from where we alternately
minimize with respect to (a) and (b). For the calculation of the
cb-norm a slightly modified version of the algorithm described in
\cite{Johnston} is used.

The Birkhoff defect shows a remarkable qualitative agreement with the
absolute volume of the tetrahedron spanned by the four Bloch vectors
$\{\vec b_1,\ldots,\vec b_4\}$ (see Fig. \ref{fig:figure3}).
Obviously, the tetrahedron volume not only enables to distinguish the
different classes of dynamics, it also gives a quantitative measure of
the quantumness of the channel.

\section{Irreversible Quantum Decoherence}
\begin{figure}[t]
  \includegraphics[width=0.9\linewidth]{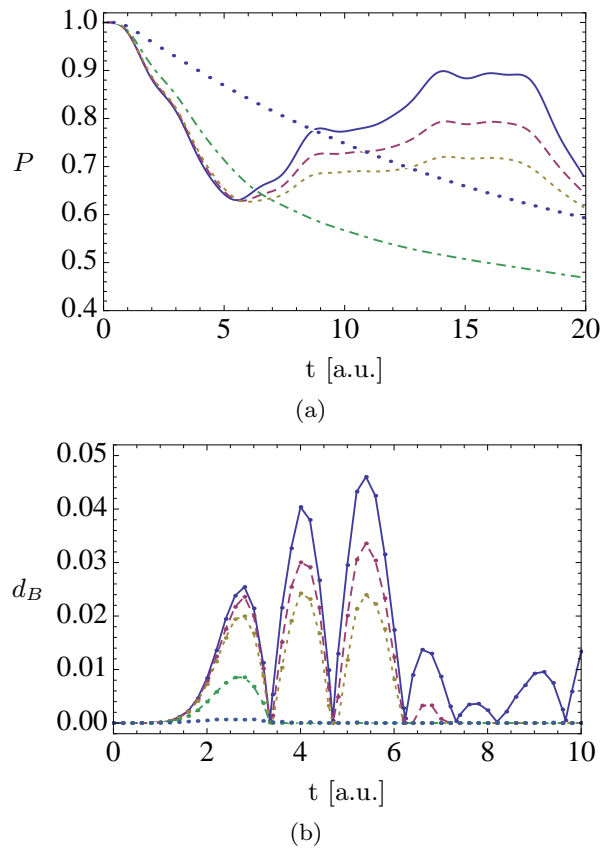}
  \caption{\small (Color online) (a) Purity $P$ of the two-qubit
    system $\sys$ and (b) Birkhoff defect $d_B$ of the quantum channel
    against time t for the single-qubit environment (solid line), and
    with additional coupling to a zero temperature bath with coupling
    strength $\gamma= 0.05 $ (long dashes)$,0.1 $ (short dashes)$,
    0.5$ (dot-dashed), and $2$ (dotted).}
  \label{fig:figure4}
\end{figure}

Clearly, the three-qubit model is fully reversible. In order to
introduce irreversibility we include an additional damping of the
reservoir qubit via spontaneous decay (cf. Fig. \ref{fig:figure1}),
leading to the Markovian master equation for the full density operator
$\rtot$ \cite{Carmichael}
\begin{eqnarray*}
  \dot\rho_{\mbox{\tiny tot}}  = -i [H, \rtot] +
  \frac{\gamma}{2} 
  \left(2 \, \sigma_{-} \rtot \sigma_{+} 
    - \sigma_{+}\sigma_{-} \rtot
    - \rtot\sigma_{+} \sigma_{-} \right), 
\end{eqnarray*}
where $H$ is the Hamiltonian describing the original model
(\ref{eq:hamiltonian}), $\sigma_\pm$ are the raising and lowering
operators acting on the reservoir qubit. The channel is again given
through the reduced dynamics $\eps_t[\rho] = \tr_\res \left(\rtot
  (t)\right) =: \rho'$. The influence of the additional damping may be
seen in the purity of the two-qubit system $\sys$ [cf. Fig.
\ref{fig:figure4} (a)].

For this extended model the FSOV criterion is no longer suitable (the
environment can obviously no longer be described as a single
qubit). Yet, the Birkhoff defect $d_B$ gives a way of examining the
nature of the channel. We observe [cf. Fig. \ref{fig:figure4} (b)]
that for increasing coupling $\gamma$ of the environment qubit to the
zero temperature bath the Birkhoff defect of an average channel
decreases until, for $\gamma$ large enough, it is zero for almost all
times. Note, however, that for small coupling $d_B$ stays well above
zero: we can still observe quantum decoherence.

\section{Conclusions}
To summarize, based on a simple model we are able to give a generic
example of a feasible two-qubit decoherence channel that does not
belong to the class of random unitary channels.  Remarkably, we see a
strong correlation between the Birkhoff defect and the volume of the
tetrahedron spanned by the Bloch vectors of the relative states of the
environment qubit. For an extension of our model including
irreversibility, we see genuine quantum decoherence as long as the
coupling to the bath is small enough.  We hope that our model will
help to further explore the difference between ``classical'' random
unitary phase damping and ``true'' quantum decoherence and thus help
to elucidate the true role of entanglement in open quantum system
decoherence.

\acknowledgements
The authors would like to thank Gernot Alber, Lajos Di\'{o}si,
Hartmut H\"affner, Florian Mintert,
Carlos Pineda,
Stephan Rietzler,
Thomas Seligman,
Markus Tiersch, and
Michael M. Wolf
for useful discussions and hints. J.H. acknowledges
support from the International Max Planck Research School Dresden. 
\bibliographystyle{apsrev}

\end{document}